# Emergent Electronic Kagome Lattice in Correlated Charge–Density–Wave State of 1$T$-TaS$_2$

Haoyu Dong[1, +], Peihan Sun[1, +], Le Lei[1,2, +], Yanyan Geng[1], Jianfeng Guo[1], Yan Li[2], Li Huang[2], Rui Xu[1], Fei Pang[1], Wei Ji[1], Weichang Zhou[3], Zheng Liu[4], Zhong-Yi Lu[1], Hong-Jun Gao[2], Kai Liu[1, *], and Zhihai Cheng[1, *]

[1]*Beijing Key Laboratory of Optoelectronic Functional Materials & Micro-nano Devices, Department of Physics, Renmin University of China, Beijing 100872, China*

[2]*Beijing National Laboratory for Condensed Matter Physics, Institute of Physics, Chinese Academy of Sciences, Beijing 100190, China*

[3]*Key Laboratory of Low-dimensional Quantum Structures and Quantum Control of Ministry of Education, College of Physics and Information Science, Hunan Normal University, Changsha 410081, China*

[4]*Institute for Advanced Study, Tsinghua University, Beijing 100084, China*

**Abstract:** Quantum materials with tunable correlated and/or topological electronic states, such as the electronic Kagome lattice, provide an ideal platform to study the exotic quantum properties. However, the real-space investigations on the correlated electronic Kagome lattice have been rarely reported. Herein, we report on the electronic Kagome lattice emerging in the correlated charge–density–wave (CDW) state of 1$T$-TaS$_2$ at ~200 K via variable-temperature scanning tunneling microscopy (VT-STM). This emergent Kagome lattice can be considered a fractional electron-filling superstructure with reduced translational and rotational symmetries, confirmed by STM measurements and density functional theory simulations. The characteristic band structure and density of states of this electronic Kagome lattice are further explored based on theoretical calculations. Our results demonstrate a self-organized electronic Kagome lattice from the correlated CDW state via the effective tuning parameter of temperature and provide a platform to directly explore the interplay of correlated electrons and topological physics.

\* Email address: zhihaicheng@ruc.edu.cn, kliu@ruc.edu.cn

**Introduction**

Novel correlated and topological electronic states are two important themes in condensed matter physics [1–3] owing to their fundamental scientific value and potential applications in future dissipation-less electronics and quantum computers. Quantum materials composed of atoms arranged on a Kagome lattice with corner-sharing triangles, such as $Co_3Sn_2S_2$ [4–6], $A$$V_3Sb_5$ ($A$ = K, Rb, Cs) [7–9], and $Cu_3Zn(OH)_6Cl_2$ [10, 11], provide fascinating platforms for exploring these novel quantum states. The topological Weyl semimetal state and the giant intrinsic anomalous Hall effect (AHE) [12–16] have been found in the magnetic Kagome metals $Co_3Sn_2S_2$ and $Fe_3Sn_2$. The recently discovered Kagome compounds $A$$V_3Sb_5$ ($A$ = K, Rb, or Cs) have been shown to exhibit rich correlated electronic states, such as the unusual charge order accompanying time-reversal symmetry breaking (TRSB) at high temperatures and exotic superconductivity at low temperatures [17]. Owing to the intrinsic geometric frustration in Kagome materials, the quantum spin liquid (QSL) state has also been proposed theoretically and has been tentatively explored experimentally in insulating Kagome antiferromagnets, among which one of the most representative materials is $Cu_3Zn(OH)_6Cl_2$ [18]. In addition to the aforementioned atom-based Kagome lattice materials, the electronic Kagome lattices would also offer a chance to explore correlated and topological states. One possible way to realize the novel electronic Kagome lattice is modulating the fractional filling of tunable correlated electronic states in triangular electron lattice materials.

Recently, the gate-tunable correlated electronic states in different fractional filling have been realized at the triangular Moiré superlattices with flat minibands in the magic-angle twisted-bilayer graphene, ABC-stacked trilayer graphene, and $WSe_2/WS_2$ heterostructures, which are considered as Mott and generalized Wigner crystal states [19–26]. The superconductivity, AHE, ferromagnetism, and other exotic states have also been discovered in these superlattice platforms at different fractional electron filings of the flat minibands via gating. For example, the generalized Wigner crystals at different fractional fillings (n = 1/3, 1/2, and so on) have been detected on both electron- and hole-doped sides by using the novel optically detected resistance and capacitance (ODRC) [24] and scanning microwave impedance microscopy (SMIM) techniques in the $WSe_2/WS_2$ hetero-bilayers [25]. The corresponding real-space electron-filling configurations at n = 1/3 (and 1/2, 2/3) are further directly visualized as commensurate triangular (and stripe, honeycomb) electron lattices by using a novel scanning tunneling microscopy (STM) scheme [26]. The Kagome electron lattice at n = 3/4 exhibits these interesting topologically correlated electronic states as reported in twisted-bilayer graphene via transport and other global methods [27–29]. However, this type of electronic Kagome lattice has not thus far been directly visualized in real space.

Another example that possesses interesting electronic super lattice is the charge–density–wave (CDW) materials. Some specific long-range-ordered CDW states with flat band can host similar correlated electronic states as twisted-bilayer graphene, such as the Mott–Hubbard state in the $\sqrt{13} \times \sqrt{13}$ commensurate CDW (CCDW) state of 1$T$-TaS$_2$, 1$T$-TaSe$_2$, and 1$T$-NbSe$_2$ at low temperatures [30–32]. Recently, it was theoretically proposed that artificial electronic lattices can be realized on the two-dimensional Mott insulator 1$T$-TaS$_2$ by means of metal adsorption [33]. The CDW superlattices are usually sensible to the charge doping, pressure, and lattice strain [34], however, they may be also tuned via the temperature-dependent electron–phonon coupling to realize some fascinating electronic states.

In this study, we observed an emergent electronic Kagome lattice in the correlated CDW states of 1$T$-TaS$_2$ via variable-temperature scanning tunneling microscopy (VT-STM) by using the temperature as a tuning parameter. The detailed atomic structure underlying this novel electronic superlattice was determined by combining high-resolution STM image and density functional theory (DFT) simulations, and its unique band structure and partial density of states (PDOS) were further explored based on theoretical calculations. The equivalent honeycomb pattern in ling graph of the emergent electronic Kagome lattice was also observed and discussed based on band- and energy-specific charge density analyses. Our studies on 1$T$-TaS$_2$ provide a convenient way to realize the electronic Kagome lattice and may also apply to other quantum materials with similar properties.

# Results

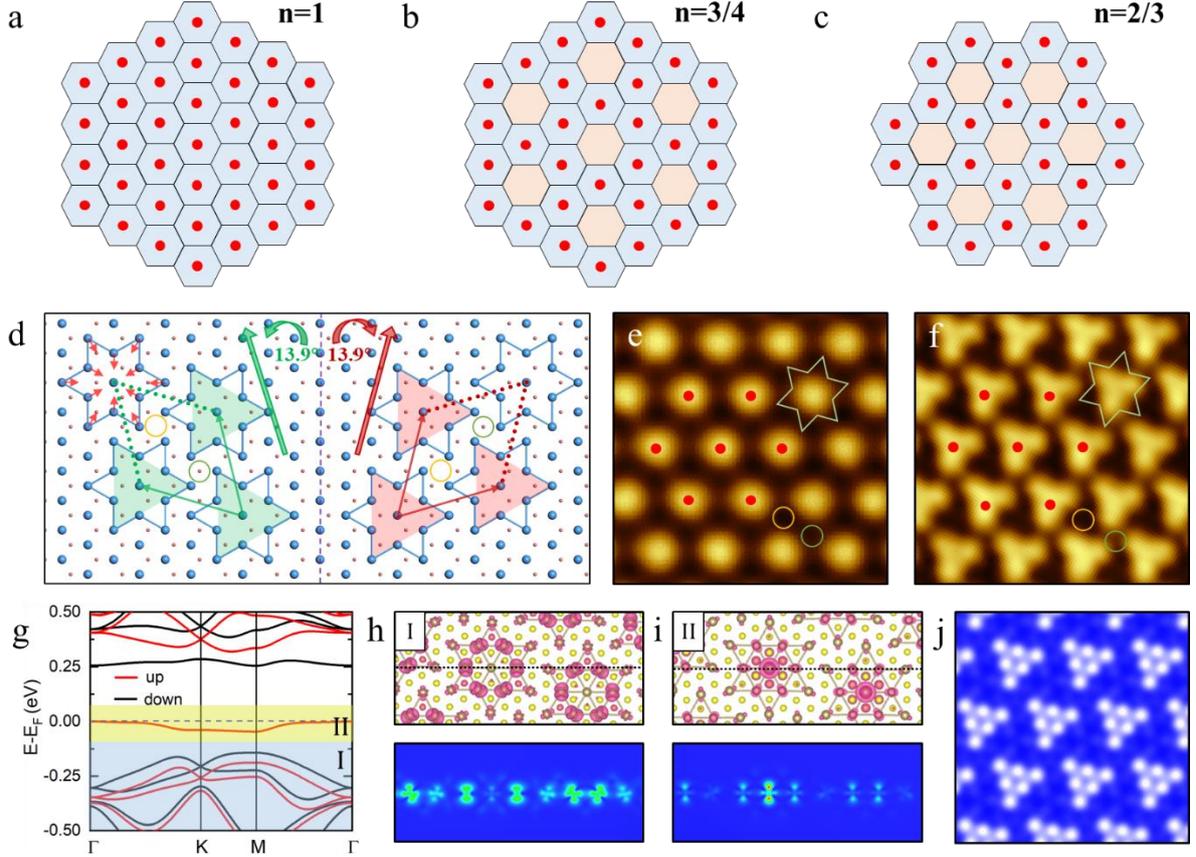

**FIG. 1. Different electron-filling patterns of the triangular lattice and the correlated charge–density–wave (CDW) state in 1$T$-TaS$_2$.** (a) Schematic of the triangular lattice filled by one electron per unit cell (n = 1). (b, c) Patterns for the triangular lattice with n = 3/4 and n = 2/3 electron fillings. (d) The correlated CDW state with the Star of David (SD) cluster in 1$T$-TaS$_2$. The blue and red balls represent the Ta and S (only top surface) atoms, respectively. The SD clusters are arranged with a √13 × √13 periodicity in the right-chiral (red) or left-chiral (green) phases. (e, f) Scanning tunneling microscopy (STM) images of 1$T$-TaS$_2$ showing the round (e, 5 × 5 nm$^2$, -0.2 V, -100 pA) and triangular (f, 5 × 5 nm$^2$, -0.5 V, -100 pA) shapes of the SDs in the correlated CDW state. Dark hollow (DH) and shallow hollow (SH) sites are marked on the models and the STM images using the orange and green circles, respectively. (g) Calculated band structure of monolayer 1$T$-TaS$_2$ in the CDW state. The red and black solid lines indicate the spin-up and spin-down bands, respectively. (h, i) Top views of the charge densities $\rho_I$ and $\rho_{II}$ integrated over the respective energy windows I and II. The isosurface values for $\rho_I$ and $\rho_{II}$ are 0.005 e/Å$^3$ and 0.0012 e/Å$^3$, respectively. The side views (in the lower panel) are drawn on the vertical planes along the dashed lines in the upper panels. (j) Simulated STM image (V = -500 mV, $\rho$ = 1.2 × 10$^{-5}$ e/Å$^3$).

We start from considering a triangular lattice model with different electron fillings. In the case of n = 1 Mott-insulating state, the entire triangular lattice is occupied uniformly with exactly one electron per unit cell [Fig. 1(a)]. In the generalized Wigner crystal states with fractional electron fillings at n = 1/3 and n = 1/4, the correlated electrons can only occupy every second (and third) nearest-neighbor (NN) sites, respectively (Fig. S1). By contrast, for those at n = 3/4 and n = 2/3 fillings, the depleted electrons, i.e., holes, occupy the respective third and second NN sites and form the Kagome lattice [Fig. 1(b)] and the honeycomb lattice [Fig. 1(c)], respectively. These fractional filling states could be described by the extended Hubbard model with nonlocal inter-site interactions. They have also been realized in the real strong correlated electron materials, such as the Moiré superlattices of twisted graphene and transition metal dichalcogenides, via the application of tunable gating voltages [19–32].

Different from the Moiré superlattices of van der Waals (vdW) materials, the layered 1$T$-TaS$_2$ provides a CDW-driven chiral triangular superlattice formed by correlated electrons, as shown in Fig. 1 (d-j). At low temperatures (below ~180 K), 1$T$-TaS$_2$ enters a CCDW ground state with periodic atomic distortions due to strong electron–phonon coupling [35]. The common building block of the CCDW phase, known as the SD cluster, is composed of 13 distorted Ta atoms and forms a $\sqrt{13} \times \sqrt{13}$ triangular superlattice [Fig. 1(d)]. It is noted that there are left- and right-chiral CDW phases (Fig. S2) because of the relative rotations (±13.9°) of the superlattice with the underlying atomic lattice. The CDW unit of SD could be selectively imaged as round and triangular protrusions [Fig. 1(e, f)] using STM depending on the tunneling bias. The chirality and two types of hollow sites in the CDW phase of 1$T$-TaS$_2$ can be determined directly using the STM image, as marked in Fig. 1(d-f).

To gain more information about the electronic structure of the CCDW phase, we performed DFT calculations on 1$T$-TaS$_2$. Each SD cluster in the CCDW phase of 1$T$-TaS$_2$ contains thirteen Ta 5$d$ electrons, among which twelve electrons pair to form six fully occupied bands and one unpaired electron is left in a half-filled band. This half-filled band is very flat because the electron hopping between neighboring SDs is limited by a large distance (~1.2 nm). After taking the electronic correlation and spin of Ta-5$d$ electrons into account, this half-filled flat band splits into the upper and lower Hubbard bands, opening a gap of ~0.3 eV for U = 2.27 eV [Fig. 1(g)]. These calculation results are consistent with those in previous reports [36–40]. The calculated total magnetic moment of each SD equals to 1 $\mu_B$, which is mainly contributed by the $5d_{z^2}$ orbital of the central Ta atom. The charge densities $\rho_I$ and $\rho_{II}$ integrated over two energy windows I ([-0.50, -0.08] eV) and II ([-0.08, 0.12] eV) are displayed in the upper panels of Figs. 1(h) and 1(i), respectively. We find that $\rho_I$ exhibits a charge distribution on the six NN Ta atoms encircling around the SD center, while $\rho_{II}$ shows a quasi-localization at the central Ta atom for which the characteristics of Ta $5d_{z^2}$ orbital can be clearly observed [bottom panel of Fig. 1(i)]. Figure 1(j) shows the simulated STM image at a bias voltage of -0.5 V, which is in good agreement with the experimental image in Fig. 1(f).

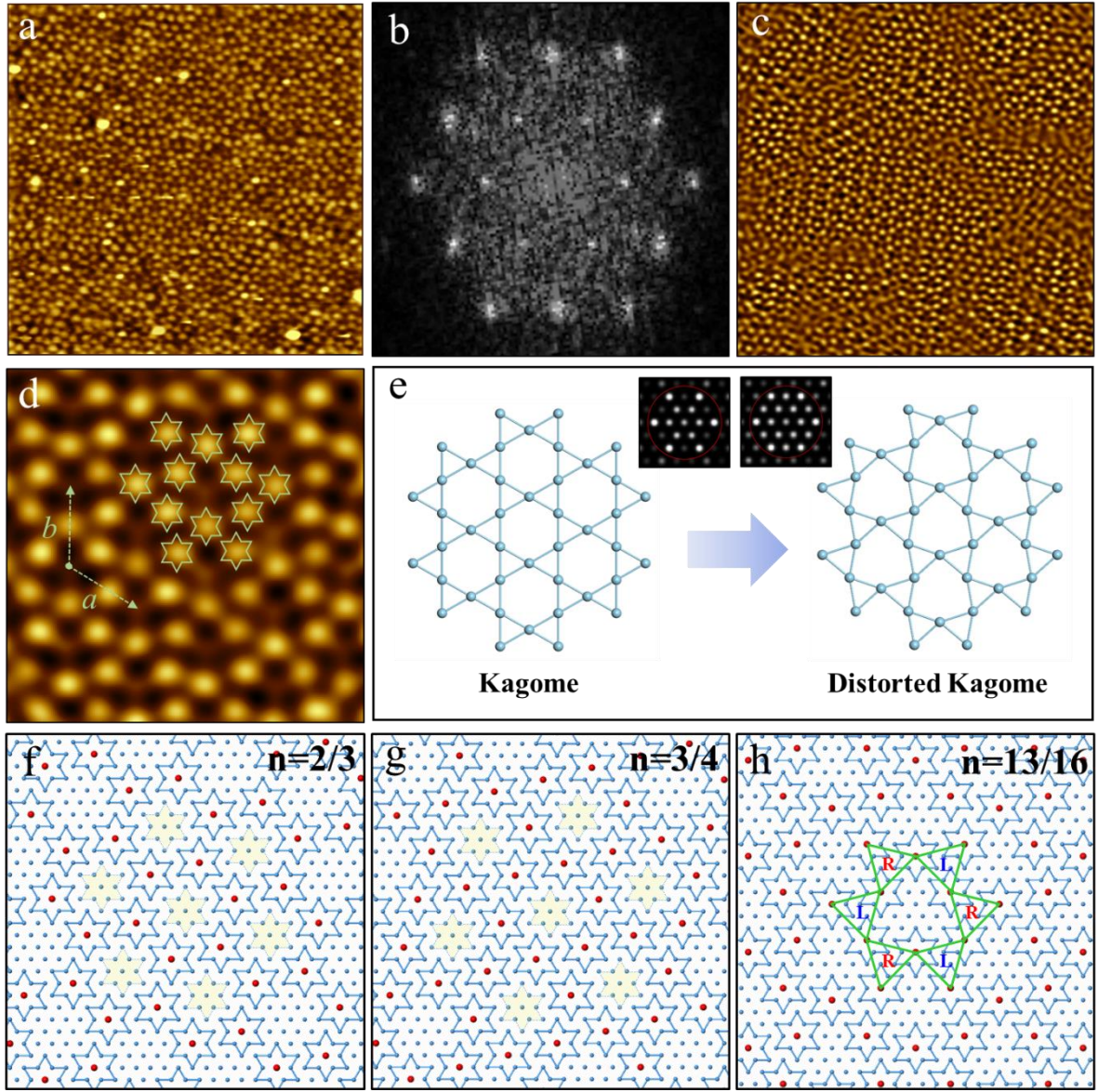

**FIG. 2. Electronic Kagome lattice arising from the partial thermal melting of the correlated CDW state (triangular lattice) in 1*T*-TaS$_2$.** (a) Large-scale STM image of the SD lattice (40 × 40 nm$^2$, -1.3 V, -300 pA). (b) The corresponding Fast Fourier Transform (FFT) of (a) shows the ordering of SDs in the STM image. (c) The FFT-filtered STM image of (a). (d) High-resolution STM image of the electronic Kagome lattice. The SDs are denoted by overlaid stars. The superlattice constants are determined to be ~2.3 nm for the primitive vectors *a* and *b* (10 × 10 nm$^2$, -1.3 V, -300 pA). (e) Schematics of the ideal and distorted Kagome lattices. The insets show the corresponding FFTs of the ideal and distorted Kagome lattices. (f, g) Schematic models for the fractional fillings of the triangular SD lattice in the correlated CDW state of 1*T*-TaS$_2$ at n = 2/3 (f) and n = 3/4 (g). (h) Proposed schematic model for the experimentally observed electronic Kagome lattice in 1*T*-TaS$_2$.

Upon increasing the temperature, the commensurate CDW phase transforms into a nearly commensurate CDW phase (NCCDW) via a first-order transition at ~200 K. According to previous works [41, 42], the stacking-dependent interlayer interactions almost disappear, while the intralayer electron correlated interaction and electron–phonon coupling play dominant roles during this hysteretic phase transition. In our study, some particular areas on $1T$-$TaS_2$ were also observed at ~200 K with the "amorphously" distributed SDs [as shown in Figs. 2(a) and S3], while the corresponding FFT image of Fig. 2(b) presents a clear six-fold symmetric pattern that is distinct from that of the CCDW triangular lattice (even including the coexisting L- and R-chiral domains) (Fig. S4). The FFT-filtered STM image of Fig. 2(a) demonstrates the locally ordered charge pattern of corner-sharing triangle patches with the remaining CCDW SDs (Fig. S5), which is different from the previously reported NCCDW, mosaic, and "hidden" states.

The observed novel charge pattern is clearly resolved in the STM image of Fig. 2(d), revealing periodically spaced round cluster protrusions. The measured in-plane lattice constant for this periodic pattern is ~2.3 nm, which is between the respective values of the $\sqrt{3} \times \sqrt{3}$ (~2.1 nm) and $2 \times 2$ (~2.4 nm) superlattice constants of the CCDW state. By overlaying the structural model on top of the STM image in Fig. 1(d), we can determine that the observed round protrusions are still the SDs; each hollow site in the middle of the three SDs is SH, while the central depressions could be assumed to be "melted" SDs (known as voids). The remaining ordered SDs form a corner-sharing distorted Kagome-like superlattice of "superatoms," as schematically shown in Fig. 2(e), confirmed by the consistency of the FFT patterns (Fig. S6). It can be simply assumed that the distorted Kagome superlattice of SDs emerged from the fractional thermal "melting" of the ground CCDW state in the equilibrium condition at low temperatures.

The honeycomb and Kagome superlattices of SD-superatoms could be formed at the n = 2/3 and n = 3/4 fractional fillings ($\sqrt{39} \times \sqrt{39}$ and $\sqrt{52} \times \sqrt{52}$ supercells) for the correlated CDW triangular lattice of $1T$-$TaS_2$, as shown by the atomic structural models in Figs. 2(f) and 2(g), with the 1/3 and 1/4 fractional fillings marked as "melted" SDs (voids, dashed stars). However, both models do not coincide with our experimental observations, so a new structural model is proposed for the distorted Kagome superlattice based on the determined parameters (Fig. S7), as shown in Fig. 2(h). In contrast to the perfect Kagome lattice with a filling of n = 3/4, the distorted SD Kagome lattice exhibits a 13/16 filling pertaining to the CDW SD superlattice.

The √48 × √48 supercell of this distorted Kagome lattice consists of 3 SDs (39 Ta atoms) and one "melted" void (9 Ta atoms). The structural parameters, including the superlattice constants and the high-symmetric elements, are fully consistent with our experimental results.

Different from the intrinsic atom-based Kagome lattice (e.g., $AV_3Sb_5$), the observed Kagome lattice in 1$T$-$TaS_2$ is a pure-electron-dominated lattice that emerged within the correlated CDW phase and is similar to the gate-tunable 3/4 fractional filling of the Moiré superlattice of vdW materials. The key difference is that the CDW superlattice and SD unit are from the periodic atomic distortions due to the electron–lattice (electron–phonon) interactions, which could be modified by the excitation of correlated electrons. The underlying mechanism for the thermal transition from the ground CDW superlattice to the emerged Kagome lattice in 1$T$-$TaS_2$ is very complicated. The translational and rotational symmetries are reduced (√13 × √13 to 4√3 × 4√3; six-fold to three-fold), while the mirror-symmetry is recovered (chirality to achirality) at the atomic structure side based on the increased thermal entropy (Fig. S8). On the electronic side, the correlated electronic states could also play important roles in the emergence of this electronic Kagome lattice.

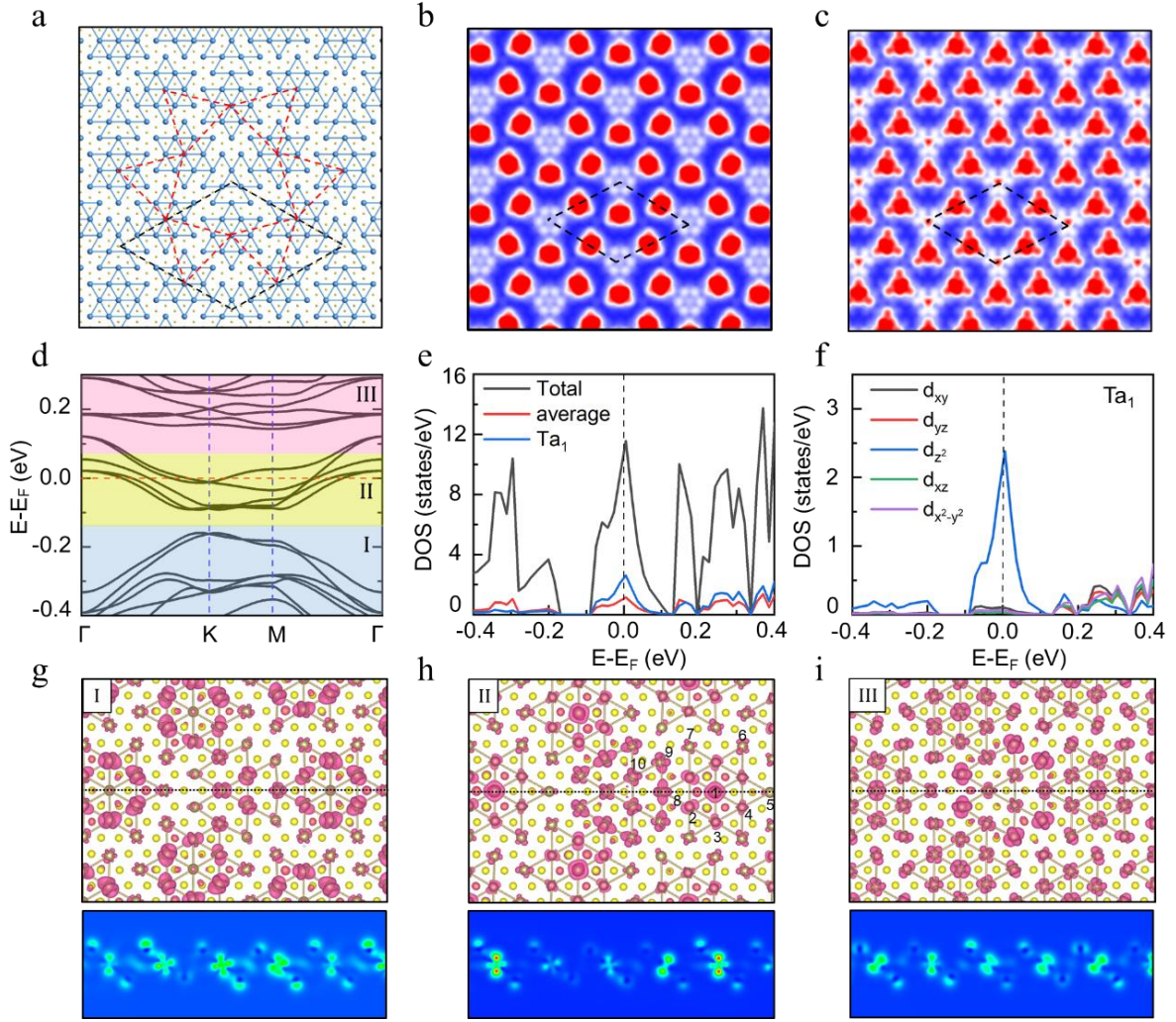

**FIG. 3. Calculated atomic and electronic structures of the electronic Kagome lattice in 1$T$-TaS$_2$ based on density functional theory (DFT).** (a) Optimized atomic structural model for the electronic Kagome lattice. (b, c) Simulated STM images for $V$ = -100 mV, $\rho = 1 \times 10^{-5}$ eÅ$^{-3}$ (b) and $V$ = -500 mV, $\rho = 1 \times 10^{-5}$ eÅ$^{-3}$ (c), respectively. (d) Calculated band structure of the electronic Kagome lattice. Three specific energy windows I-III are marked with different colors. (e) Total density of states (DOS), average DOS of all Ta atoms, and local DOS of the central Ta atom (Ta$_1$) in the SD unit. The labeling of Ta atoms is shown in (h). (f) Partial DOS of Ta$_1$ atom. (g–i) Top views of the charge densities $\rho_I$, $\rho_{II}$, and $\rho_{III}$ integrated over the three energy windows I-III (d), respectively. The isosurface value of $\rho$ is set to 0.00275 e/Å$^3$. The side views (in lower panels) are drawn on the vertical planes along the dashed lines in the corresponding upper panels.

To identify the detailed atomic structure of the electronic Kagome lattice of 1$T$-TaS$_2$, we also performed DFT calculations based on the proposed structural model in Fig. 2(h). The fully relaxed structure is shown in Fig. 3(a), which consists of three SD units and nine bonded Ta atoms at the central "melted" void. The simulated STM images at different bias voltages are displayed in Figs. 3(b), 3(c), and Fig. S9, being in good agreement with our experimental results. Among them, the simulated STM images of SD are close to a circle at -0.1 V and a triangle at -0.5 V, which are similar to the patterns of SD in the CCDW state as shown in Figs. 1(e) and 1(f). In addition, we also tried another possible structure with the central 9 Ta atoms fixed in their undistorted positions but found that its total energy was 211.3 meV/cell higher than that of above optimized structure [Fig. 3(a)]. Hence, we focus on the electronic properties of the optimized structure in the following.

Figure 3(d) shows the electronic band structure of the electronic Kagome lattice of 1$T$-TaS$_2$ along the high-symmetry path of Brillouin zone (BZ) calculated without the spin–orbital coupling (SOC). There are five bands passing through the Fermi level $E_F$, justifying the metallic behavior. Based on the number of counted electrons, these five bands are occupied by six electrons, three of which come from the three SD units, and the other three are provided by the bonded Ta atoms at the central "melted" voids between the SDs [Fig. 3(a)]. As there is hybridization between the Kagome lattice of SD and the central "melted" Ta atoms through the Ta$_8$ atoms [Figs. 3(h) and S10(c)], the characteristics of typical Kagome bands, namely two linearly crossed bands at K point and a very flat band across the BZ, are perturbed to some extent. In Fig. 3(e), the narrow-band nature of these five bands results in a sharp peak in the total density of states (DOS) around $E_F$. Unlike the results of the ground CCDW phase, this peak exhibits a certain degree of broadening owing to the band dispersion. According to the atomic- and orbital-resolved DOSs displayed in Figs. 3(e), 3(f), and S10, we can infer that the electronic states around the $E_F$ are mainly contributed by Ta 5$d$ orbitals, among which the $d_{z^2}$ orbital of the Ta$_1$ atom located in the SD center dominates. We also examined the electronic structure with the inclusion of SOC and found that the energy dispersion was almost unchanged (Fig. S11). Hence, we will not consider the SOC effects further.

To gain a deeper understanding of the electronic properties of this correlated electronic Kagome lattice, we plotted the real-space distributions of the electronic states within three different energy windows [as marked in Fig. 3(d)]. Specifically, the electronic states $\rho_I$ [Fig. 3(g)] located in energy window I ([-0.4, -0.1] eV) are mainly distributed on the six Ta atoms around the SD center and the three Ta atoms in between the SDs. In the energy window II ([-0.1, 0.14] eV), the central Ta$_1$ atoms of SDs play a dominant role in electronic states $\rho_{II}$ followed by the $Ta_8$, $Ta_9$, and $Ta_{10}$ atoms in between the SDs [Fig. 3(h)]. Moreover, the dominant orbital characteristics of $d_{z^2}$ on $Ta_1$ atoms are clearly visible in the lower panel of Fig. 3(h). Regarding $\rho_{III}$ in the energy window III ([0.14, 0.3] eV), the $Ta_9$ and $Ta_1$ atoms have larger contributions than other atoms [Fig. 3(i)]. Overall, the SDs in this distorted electronic Kagome lattice [Figs. 3 and 2(h)] have stronger interactions than those in the CCDW state (Fig. 1).

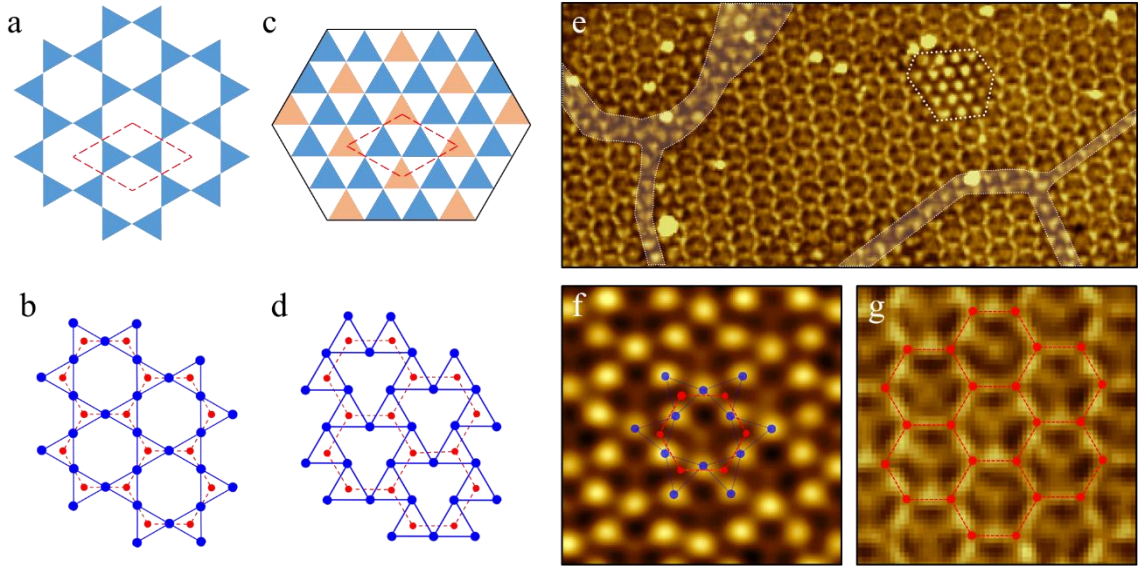

**FIG. 4. Kagome and honeycomb patterns observed on 1$T$-TaS$_2$.** (a, b) The Kagome lattice (a), and the line graph (b) of a Kagome lattice constructed from a parent honeycomb lattice. (c, d) Coloring-triangle (CT) lattice (c), and the line graph (d) of a CT lattice constructed from a parent honeycomb lattice. (e) Large-scale STM image of the observed honeycomb patterns with the remaining correlated CDW patches, and domain boundaries (55 × 25 nm$^2$). (f, g) High-resolution STM image of the observed Kagome (f, 9 × 9 nm$^2$) and honeycomb (g, 9 × 9 nm$^2$) patterns with the overlaid schematics of the Kagome and honeycomb lattices. The inset shows the structure of Kagome (blue) and the honeycomb lattices (red).

The correlated electronic Kagome lattice provides an interesting and flexible platform to study the interplay between frustrated geometry, correlation, topology, and electron–lattice interactions. The band structures of the Kagome lattice consist of the topological flat and Dirac bands, while the location of the flat band depends on the inter-site hopping integral. Recently, Liu *et al.* discovered a triangular lattice type known as the coloring-triangle (CT) lattice, which is equivalent to the Kagome lattice mathematically following the application of a unitary transformation and line-graph construction, and contains the identical Kagome bands as that of the Kagome lattice [43, 44]. As illustrated in Figs. 4(a) and 4(b), the Kagome lattice can be realized by blocking the hopping around one lattice site in a 2 × 2 supercell of a triangular lattice, while the CT lattice can be realized by blocking the hopping around the center of a triangle in a √3 × √3 supercell of a triangular lattice [Figs. 4(c) and 4(d)]. The observed distorted electronic Kagome lattice in 1$T$-TaS$_2$ is between the Kagome and CT lattices, and also contains a line graph of a honeycomb lattice.

The apparent honeycomb pattern in the electronic Kagome state of 1$T$-TaS$_2$ has also been observed in our experiments depending on the tunneling conditions (Fig. S13). Figure 4(e) shows a large-scale STM image of honeycomb patches with the remaining CDW SD regions as the domain boundaries and the coexisting small islands. Figures 4(f) and 4(g) show the high-resolution STM images of the observed Kagome and honeycomb patterns with the overlaid schematic site-bonded lattice models. The underlying mechanism for the observed honeycomb pattern could be the selective tunneling of the electronic Kagome states. The band-specific and energy-specific charge density mappings are calculated and shown in Figs. S14 and S15. As indicated, the electronic states of bands 1, 2 and some energy range represent a rough honeycomb pattern, thus indicating that the emergence of a honeycomb-like pattern in STM image is theoretically reasonable.

**Discussion**

The underlying mechanism for the formation of distorted electronic Kagome lattice in the correlated CDW state of 1$T$-TaS$_2$ at ~200 K is very complicated and only phenomenally discussed in this study. At this critical temperature, 1$T$-TaS$_2$ underwent a transition from an interlayer stacking order with strong interlayer dimerization (at low temperatures) to the one with dominant intralayer correlated electron interactions [41, 42]. Another possible effect of temperature at ~200 K is the thermal-entropy-driven weakening of the underlying CCDW lattice in 1$T$-TaS$_2$. In general, the formation mechanism can be attributed to the delicate interplay between the electron–electron correlated interactions and the electron–phonon coupling during the first-order CDW transition of 1$T$-TaS$_2$. The unambiguous formation mechanism needs to be studied theoretically further, which is beyond the scope of this study. We also notice that similar super-modulations were observed in the correlated CDW states of 1$T$-TaS$_2$ (or 1$T$-NbSe$_2$, 1$T$-TaSe$_2$) (Fig. S16) [36–38, 45]. For the super-modulation (between $\sqrt{3} \times \sqrt{3}$ and $2 \times 2$) of monolayer 1$T$-TaSe$_2$, the underlying mechanism was considered as the spinon Fermi surface instability of monolayer 1$T$-TaSe$_2$, which is proposed to be a gapless QSL candidate [45]. For the $\sqrt{3} \times \sqrt{3}$ super-modulation on the K-adsorbed surface layer, it has been attributed to the dipole–dipole interaction of the adsorbates [36, 37].

The emergent electronic Kagome lattice in the correlated CDW state of 1$T$-TaS$_2$ is promising and can enable further exploration of its potential exotic quantum states with more experimental and theoretical methods: (1) The potential topological correlated electronic characteristics can be investigated via transport and other global methods. (2) The spin configurations of electronic Kagome lattice can be explored by spin-polarized STM. (3) The coexistence and interplay among the local electrons/spin in the remnant DSs and the itinerant electrons in the melted DSs of the frustrated Kagome structure is also an interesting topic that calls for deeper theoretical investigations beyond the DFT calculations. (4) The self-organized behavior of correlated electrons also inspires the possible artificial construction of electronic lattices via the controlled electron manipulations. Recently, the directive real-space investigations of correlated electron interactions and the tentative electron manipulation have been preliminarily performed in the hole-doped correlated CDW state of the Ti-doped 1$T$-TaS$_2$.

In summary, we experimentally realized a self-organized electronic Kagome lattice in the correlated CDW state of a triangular lattice material 1$T$-TaS$_2$ at ~200 K, which can be attributed to the delicate interplay of electron-electron interaction and electron–phonon coupling during the first-order CDW transition of 1$T$-TaS$_2$. Compared with the twisting method or the atom adsorption proposal, here the temperature is used as an effective and convenient tuning parameter to introduce the Kagome pattern. The detailed atomic and electronic structures of this novel electronic superlattice was studied by combining STM measurements and DFT simulations. This emergent correlated electronic Kagome lattice provides a unique platform to directly explore the exotic quantum states via the interplay of correlated electrons and topological physics in real space and has potential applications in future quantum devices.

## Materials and Methods
### Sample preparations and scanning tunneling microscopy measurements

High-quality 1$T$-TaS$_2$ single crystals supplied by HQ Graphene were grown using a chemical vapor transport (CVT) method with iodine as a transport agent. The samples were cleaved at room temperature in an ultrahigh vacuum at the base pressure of $2 \times 10^{-10}$ Torr and were quickly transferred to the VT-STM system (PanScan Freedom, RHK, USA). VT-STM measurements were realized by using liquid-He-free cryocooler technology with adjustable heating power at the cold head. Chemically etched tungsten tips were used for constant-current STM measurements. The tips were prepared and calibrated on Ag (111) (MaTeck, Germany) surface cleaned by repeated cycles of argon ion sputtering and annealing at 550 K. The STM measurements were mostly performed at the temperature of ~200 K. At this temperature, the 1$T$-TaS$_2$ system had undergone a transition from a band insulator with interlayer dimerization (the ground state at low temperatures) to the Mott-insulating state with the dominant intralayer electron interactions. The scanning tunneling spectroscopy (STS) measurements were also performed, and only the similar rough "V-shaped" features were observed, which were attributed to the thermal-broadening effect at ~200 K. The STM data were analyzed and processed with the Gwyddion software.

### DFT calculations

The DFT calculations were performed with the projector-augmented-wave method [46, 47] as implemented the Vienna ab initio simulation package (VASP) [48, 49]. The generalized gradient approximation of the Perdew–Burke–Ernzerhof (PBE) type [50] was adopted for the exchange–correlation functional. The kinetic energy cut-off of the plane-wave basis was set to 340 eV. A single layer was used to simulate the surface layer of 1$T$-TaS$_2$. For the CCDW phase and the correlated electronic Kagome lattice of monolayer 1$T$-TaS$_2$, $6 \times 6 \times 1$ and $3 \times 3 \times 1$ Γ-centered Monkhorst–Pack $k$-point meshes were adopted for Brillouin zone (BZ) sampling. The Gaussian smearing method with a width of 0.05 eV was used for Fermi surface broadening. Both lattice constants and internal atomic positions were fully optimized until the forces on all atoms were smaller than 0.01 eV/Å. A vacuum layer larger than 16 Å was utilized to eliminate the interactions between successive image slabs along the (001) direction. In the calculation of the magnetic properties of the CCDW state, an on-site Hubbard U was included for Ta $5d$ orbitals to account for the strong Coulomb repulsion. The estimated U value of 2.27 eV based on the linear-response method [51] can reproduce the experimental Mott gap of 1$T$-TaS$_2$.

## Acknowledgements

This project is supported by the National Natural Science Foundation of China (NSFC) (No. 21622304, 6167404, 61888102, 12174443, 11934020), the Beijing Natural Science Foundation (No. Z200005), the Ministry of Science and Technology (MOST) of China (No. 2016YFA0200700, 2022YFA1403103, 2019YFA0308603), and the Strategic Priority Research Program of Chinese Academy of Sciences (CAS) (No. XDB30000000, YSBR-003). Z. H. Cheng was supported by the Fundamental Research Funds for the Central Universities and by the Research Funds of Renmin University of China (No. 21XNLG27). H. Y. Dong was supported by the Fundamental Research Funds for the Central Universities and by the Research Funds of Renmin University of China (No. 22XNH095). Computational resources were provided by the Physical Laboratory of High-performance Computing at the Renmin University of China and Beijing Super Cloud Computing Center.

## Author Contributions

Z.H.C., K.L., H.-J.G. and Z.-Y.L. coordinated the research project. H.Y.D., L.L., Y.Y.G., J.F.G., S.M., Y.L., L.H. and F.P. performed STM experiments. W.C.Z. prepared samples. P.H.S., W.J., Z.L. carried out the DFT calculations. All of the authors participated in analyzing experimental data, plotting figures, and writing the manuscript.

## Competing Interests

The authors declare no competing financial interests.

## Data Availability

The authors declare that the data supporting the findings of this study are available within the article and its Supplementary Information.